\documentclass[12pt,a4paper]{article}

\RequirePackage[l2tabu, orthodox]{nag}

\usepackage{mjheppub}
\usepackage{color}
 \usepackage{comment}
\usepackage{lipsum}
\usepackage{amsthm,amssymb,amsmath,epic,eepic,float}
\usepackage{rotating,epsfig,indentfirst,array,varioref}
\usepackage{appendix,marginnote,tikz,pgf,mathtools}
\usepackage{setspace}
\usepackage{mjheppub}

\usepackage{tikz}
\usetikzlibrary{decorations.pathreplacing}
\usetikzlibrary{calc}

\usepackage{color}

\def\leq{\leqslant}

\newcommand{\bal}{\begin{equation}\begin{aligned}}
\newcommand{\eal}{\end{aligned}\end{equation}}
\newcommand{\RR}{\mathbb{R}}
\newcommand{\ch}{\mathrm{ch}}
\newcommand{\co}{\mathrm{cos}}
\newcommand{\si}{\mathrm{sin}}
\newcommand{\sh}{\mathrm{sh}}

\newcommand{\nn}{\nonumber}

\newdimen\tableauside\tableauside=1.0ex
\newdimen\tableaurule\tableaurule=0.4pt
\newdimen\tableaustep
\def\phantomhrule#1{\hbox{\vbox to0pt{\hrule height\tableaurule
width#1\vss}}}
\def\phantomvrule#1{\vbox{\hbox to0pt{\vrule width\tableaurule
height#1\hss}}}
\def\sqr{\vbox{%
  \phantomhrule\tableaustep

\hbox{\phantomvrule\tableaustep\kern\tableaustep\phantomvrule\tableaustep}%
  \hbox{\vbox{\phantomhrule\tableauside}\kern-\tableaurule}}}
\def\squares#1{\hbox{\count0=#1\noindent\loop\sqr
  \advance\count0 by-1 \ifnum\count0>0\repeat}}
\def\tableau#1{\vcenter{\offinterlineskip
  \tableaustep=\tableauside\advance\tableaustep by-\tableaurule
  \kern\normallineskip\hbox
    {\kern\normallineskip\vbox
      {\gettableau#1 0 }%
     \kern\normallineskip\kern\tableaurule}%
  \kern\normallineskip\kern\tableaurule}}
\def\gettableau#1 {\ifnum#1=0\let\next=\null\else
  \squares{#1}\let\next=\gettableau\fi\next}

\def\be{\begin{equation}}
\def\ee{\end{equation}}
\def\ba{\begin{array}}

\def\d{\partial}
\def\dd{\mathrm{d}}

\def\nn{\nonumber}

\restylefloat{figure}

\renewcommand{\d}{\partial}

\newcommand{\FF}{\mathcal{F}}
\newcommand{\UU}{\mathbb{U}}
\newcommand{\WW}{\mathbb{W}}

\def\nn{\nonumber}

\def\pd #1{\frac{\partial}{\partial #1}}

\def\Res{\operatorname{Res}}

\begin{document}

\title{\boldmath 
Minimal gravity and Frobenius manifolds: bulk correlation on sphere and disk
}

\author[a]{Konstantin Aleshkin,}
\author[b\,\dagger]{Vladimir Belavin,}
\note[$\dagger$]{Weston Visiting Professorship at Weizmann Institute. On leave from
 Lebedev Physical Institute, Moscow.}
\author[c]{Chaiho Rim\,}

\affiliation[a]{L.D. Landau Institute for Theoretical Physics, Akademika Semenova av., 1-A, \\
Chernogolovka, 142432  Moscow region, Russia,
 \newline
 International School for Advanced Studies (SISSA), via Bonomea 265, 34136 Trieste, Italy}
\affiliation[b]{\vspace{2mm} 
Department of Particle Physics and Astrophysics, Weizmann Institute of Science,\\
Rehovot 7610001, Israel}
\affiliation[c]{\vspace{2mm}
Department of Physics, Sogang University, Seoul 121-742, Korea}

\emailAdd{kkcnst@gmail.com,
belavin@lpi.ru,
rimpine@sogang.ac.kr
}

\abstract{
There are two alternative approaches to the minimal gravity -- direct 
Liouville approach and matrix models. Recently there has been a certain progress
in the matrix model approach, growing out of presence of a
Frobenius manifold (FM) structure embedded  in the theory. The previous studies were 
mainly focused on the spherical topology. Essentially, it was shown that the action principle of 
Douglas equation allows to define the free energy and to compute the correlation numbers
if the resonance transformations are properly incorporated. The FM structure allows to find the explicit
form of the resonance transformation as well as the closed expression for the partition function.  In this paper we  elaborate on the case of gravitating disk.  
We focus on the bulk correlators and show that in the similar way as in the closed topology 
the generating function can be formulated using the set of flat coordinates on the corresponding FM. Moreover, 
the resonance transformations, which follow from the spherical topology consideration, are exactly those needed to 
reproduce FZZ result of the Liouville gravity approach.
}

\keywords{
 Conformal field theory, 2-dimensional gravity, non-critical string theory
}

\maketitle
\flushbottom

\section{Introduction} \label{sec:1}

The conformal bootstrap solution of Liouville field theory (LFT)~\cite{Polyakov1, BPZ, DO, LFT} leads to the possibility\footnote{For recent development aimed to give a rigorous proof of the LFT construction and also of its integrable structure, see~\cite{David:2014aha,Kupiainen:2017eaw}.} to formulate the ``direct''  path integral approach to $c < 1$ {\it minimal string theory}, known also as two-dimensional {\it minimal  Liouville gravity} (MLG)~\cite{Polyakov1}. 
Computations in this framework are rather complicated, since in all but simplest cases they require separate analysis of the correlation functions in both gravitational (LFT) and matter (Minimal Model) sectors~\cite{BZ1}, a careful treatment of the discrete terms~\cite{Aleshkin:2016snp} arising in the  Liouville operator product expansion, integrating over moduli spaces of curves, etc. On the other hand, the ``dual'' matrix models (MM) approach~\cite{Kazakov:1985ea,Kazakov:1986hu, KPZ}, based on the discretization of the path integral and consequent double scaling limit, provides an efficient alternative description of 2D gravity and also reveals an integrable structure of the theory through the connection with a certain class of integrable hierarchies~\cite{Douglas:1989dd, Witten:1990hr,Dijkgraaf:1990rs, Kontsevich1992}. Different checks in 90s (see, e.g.~\cite{Kazakov:1989bc, Staudacher:1989fy, Douglas:1989ve,  Gross:1989vs, Brezin:1990rb,  MSS}) pointed out that matrix models  are connected with the minimal Liouville gravity if the connection is properly understood. More elaborate checks performed on the level of correlation functions confirmed this assumption, after the discovery of Liouville  higher equations of motion~\cite{HEM} opened the way of analytic computations of the moduli integrals~\cite{Belavin:2006ex}. In particular, for the minimal gravity models of Lee-Yang series $M(2/p)$, corresponding to one-matrix models, the explicit form of the resonance transformations  form KdV time-parameters of MM to the Liouville coupling constants of MLG has been established in~\cite{Belavin:2008kv}. 
Further progress~\cite{Belavin:2013nba,Belavin:2014cua, Belavin:2014hsa,Belavin:2014xya,Belavin:2015ffa} has been made after establishing the connection between the Douglas string equation approach~\cite{Douglas:1989dd} to the matrix models and the Frobenius
 manifolds~\cite{Dubrovin:1992dz}. This connection allowed to find explicitly the generating function of the correlation numbers for the general case of   gravitational $M(q/p)$ minimal models. However, the consideration has been so far restricted to the case of gravitational models defined on closed Riemann-surfaces such as the sphere or the torus~\cite{Tarnopolsky:2009ec,Belavin:2010bs,Belavin:2010sr,Spodyneiko:2015xpa}, relevant for the sector of closed strings.

In this paper, we are interested in the boundary minimal Liouville gravity (BMLG), relevant for {\it open minimal strings}.  The worldsheet approach to BMLG requires analyzing boundary minimal models, LFT and ghosts and seems even more complicated compared to the closed case. In addition to the bulk MLG content, the classification of physical fields, or BRST cohomologies, in BMLG is specified  by a set of boundary changing fields leading to admissible boundary conditions \cite{Cardy:1984bb}. Constructing correlation numbers, even before taking moduli integrals, requires knowledge of the
structure constants appearing in the operator product expansion of two bulk fields, the structure constants of boundary fields, the couplings between bulk and boundary fields  and the one point functions of the identity operator for different boundary conditions~\cite{Cardy:1986gw,Cardy:1991tv,Runkel:1998he,Fateev:2000ik,Ponsot:2001ng}.

In the dual  approach the boundary effect was first considered in ~\cite{MSS}. Since then, it has been investigated in many different contexts. One may refer to some of the previous studies in RSOS models and O(N) fluctuating models~\cite{Kostov:2003uh, Bourgine:2009zt,Bourgine:2008pg}, in loop gas models \cite{Kostov:2002uq, Jacobsen:2006bn} and in (one-) \cite{Ishiki:2010wb} and (two-) \cite{Martinec:1991ht,Hosomichi:2008th}  matrix models. The boundary effect for the bulk correlation in the Lee-Yang series of the minimal gravity models was augmented by the resonance transformation in~\cite{Belavin:2010ba}. However, as mentioned above, the Lee-Yang series is represented by one matrix models and it turns out that the related FM is trivial (i.e. one-dimensional). Essentially, it means that the MM
partition function  found in~\cite{MSS} is easily translated to the MLG partition function, the only care is needed to properly take  into account the resonance transformations. Other minimal models are given in terms of  multi-matrix models, the corresponding Frobenius manifolds are multi-dimensional and higher Gelfand-Dickey hierarchies appear in general case~\cite{Belavin:2013nba,Belavin:2014hsa}. The analysis of the boundary effect on the MLG models related to  multi-dimensional Frobenius manifolds is still missing. In particular, there is no closed expression for the BMLG generating function of bulk correlation numbers available yet. In this work, we extend the analysis of Frobenius manifolds  for the bulk correlation on the disk.

The paper is organized as follows. In section~\ref{sec:2}, we recall the dual approach to MLG and provide the framework of finding bulk correlation numbers on the sphere and on the disk. It turns out that the role of the Douglas equation on the disk is slightly different from that on the sphere, where explicit dependence of the Frobenius (flat) coordinates on the Liouville couplings is required to be determined. On the other hand, the resonance transformation on the disk is the same as in the spherical topology. Even though this statement is not completely obvious,\footnote{The physical origin of the resonance transformations is the ambiguities coming from contact terms in the Euclidean correlators at the coincident points of the operators~\cite{MSS,BZ_MM}.}  it is anticipated because the  boundary operators that could additionally contribute to the coupling mixing are not considered here~\cite{Bourgine:2011cm}. In section~\ref{sec:3}, we 
consider the  effect of presence of nontrivial  Frobenius manifolds for the unitary series.  We present
 an Ising model case $M(3/4)$ as an example. 
In section~\ref{sec:4} we discuss  non-unitary case. We start with the Lee-Yang series. Even though the Frobenius manifold is one-dimensional, the uniform FM description allows to represent the disk partition function as an integral over the (flat) coordinate and hence this case fits well into the general pattern. Then we consider  $M(3/5)$ non-unitary  model. Similarly to  unitary $M(3/4)$ model, this model is based on two-dimensional Frobenius manifold. However, it turns out that the details of the construction are very different: compared to $M(3/4)$ case, where  the cosmological constant and the parameter $x$ in the Douglas equation have the same gravitational scaling dimensions, the gravitational scaling dimension in  $M(3/5)$ of theses two parameters are different. This fact leads to an uncertainty in the choice of the integration contour in the flat coordinate space. Nevertheless, carefully employing the Douglas string equation together with the resonance transformations and with a conjectural choice of the contour allows to match this case with the FZZ Liouville results~\cite{Fateev:2000ik} as well. In section~\ref{sec:concl}, we discuss some future perspectives. Some relevant formulae are listed in the appendix~\ref{sec:app}.

\section{Dual approach to MLG}
\label{sec:2}
 
 In this section we summarize the basic elements 
 of the dual approach to MLG
 and define a framework of computing 2D minimal gravity generating functions. 

\subsection{Douglas equation and Frobenius manifolds} 
\label{sec:2_frobenius}

The basic element  of the matrix models approach (in the continuum limit) is $Q$ differential operator  
 (for reviews, see~\cite{FranGins,DiFrancesco:1993cyw}). 
In $A_{q-1}$ model it contains $q-1$ variables $u_i$ 
and is represented as 
\begin{equation} 
Q = \d^{q} + \sum_{i=1}^{q-1} u_i(x) \d^{q-1-i}\;.
\end{equation}

The set $u=\{u_i\}$ is assumed to be a function of variable $x$,
representing the continuum limit of the discrete state enumeration  
in the basis of orthogonal polynomials.\footnote{For simplicity, we omit here the dependence on the parameter responsible for the genus expansion. More detailed description can be found e.g.  in the above mentioned references.} 
The functional dependence of $u$ on $x$ at $p$-critical point is established
by introducing $P= Q^{p/q} $ and requiring
\begin{equation}\label{Douglas-eqv}
[P, Q] = 1\;.
\end{equation}
This constraint, so-called Douglas equation~\cite{Douglas:1989dd}, in general reduces to a higher order non-linear differential equations.

On the other hand, the parameters $u_i$ can be considered as coordinates on a Frobenius manifold $\mathcal{M}$ with dim$_{\mathcal{M}}=q-1$. The main property of  $\mathcal{M}$ is the
structure of a commutative and associative algebra with unity (to be specified below) defined in the tangent space at each point $u$ and compatible  in a certain way  \cite{Dubrovin:1992dz} with the Riemannian structure of $\mathcal{M}$. 
By definition, Frobenius manifold is flat, so that there exist {\it flat coordinates}  $v_i$, in which the metric $\eta_{lk}$  on $\mathcal{M}$ is constant. For our purposes it will be convenient to use the flat basis and its  tangent  $e_i$.
The multiplicative rule for the elements of the tangent basis  is 
\be
e_i \,  e_j = c_{ij}^k~ e_k\;,
\ee
where in our convention $e_1 =1$, so that $c_{1j}^k= \delta_j^k$.
Using $\eta_{lk}$ one may perform raising and lowering indices, e.g. $c_{ijk} = c^l_{ij}\eta_{lk}$.
Defining property of $\mathcal{M}$ is that the structure constant $ c_{ijk} $ 
is fully symmetric in the index permutation and obeys the following constraint
 \be
 \partial _\ell \,  c_{ijk}  =  \ \partial_k  \, c_{ij\ell}\;,
 \ee
or, equivalently, $ c_{ijk} = \partial_i \partial_j \partial_k F $, where $F$ is the so-called {\it prepotential}.

The  flat coordinates  $v$ can be constructed  
in terms of the coordinates $u$ using the following explicit form of the flat metric (see, e.g. \cite{Belavin:2013nba, Belavin:2014hsa})
\be
\eta_{ij} = - q \Res_{y= \infty} \frac{e_i e_j } { Q'(y, u)} 
 = \delta_{i +j, q}\;.
\ee
Here polynomial $Q(y, u) = y^{q} + \sum^{q-1}_{i=1} u_i y^{q-1-i}$ 
($\d$  is replaced by a commuting number $y$), 
 $e_i = \partial Q(y, u)/\partial v^i$ and $Q'(y, u)= \partial Q(y, u)/\partial y$. 
It gives the following expression for the flat coordinates  $v_i = \theta_{i, 0}=\eta_{ij}v^j$, where\footnote{We introduce here more general object $\theta_{i, k}$, which will be useful in further consideration.} 
\be
\label{eq:theta_ik}
\theta_{i, k} = - \frac{\Gamma(i/q)}{\Gamma(i/q+k+1)} 
 \Res_{y = \infty} Q^{k + i/q}(y) \;,
\ee
with  non-negative integers $k$ and $1 \le i \le q-1$.
The structure constants are also given explicitly 
\be
c_{ijk} := - q \Res_{y= \infty} \frac{e_i e_j e_k } { Q'(y, u)}
 = \eta_{kl} ~c^l_{ij}\;.
\ee
We note that  $c_{1ij} = \eta_{ij}$ since $e_1 =1$. 

\subsection{Scaling dimensions and 2D  gravity}
 \label{sec:2_scaling}

Conformal field theoretical approach to 2D gravity emerged from the path integral formulation
 ~\cite{Polyakov1}. Liouville gravity 
is constructed as a direct product of three CFT's:
\be
\text{matter CFT} \times \text{LFT} \times (b,c) \text{ ghosts}\;,
\ee
with central charges $c_M, \; c_L $ and $c_{gh} = -26$ respectively.
The consistency of the theory requires the total central charge  to vanish 
\be\label{c-balance}
c_L + c_M +c_{gh} =0\;.
\ee
In fact, this condition guarantees
that the theory admits a nilpotent BRST operator. 

In the minimal Liouville gravity the matter sector is represented by $M(q/p)$ minimal model  (with $q$ and $p$ coprime, $q<p$).
The Liouville central charge $c_L = 1 + 6 Q_L^2$, where the background charge  $Q_L=b + 1/b $ and $b$ is the Liouville coupling constant, the minimal model cenrtral charge $c_M$ is conveniently parametrized as 
 $c_M = 1 - 6 q_M^2$ with $q_M=1/\beta- \beta$ and  $\beta = \sqrt{q/p}$
or explicitly $c_M = 1 - 6(p-q)^2/qp $.
The total central charge balance condition~\eqref{c-balance} then requires $b=\beta$. 

In the minimal Liouville gravity
the field content of the matter minimal model  
is coupled to the Liouville vertex operators in the gravitational sector, such that the resulting 
``dressed'' operator is BRST invarant. We consider the following BRST-invariant operators 
\be\label{phys-fields}
\WW_{m,n}  = c \cdot \bar{c} \cdot  \UU_{m,n}\;, ~~~
\widetilde{\WW}_{m,n} = \int \dd^2 z\; \UU_{m,n}\;,
\ee
where $\UU_{m,n} = \Phi_{m,n} \cdot V_{a_L}$ is constructed from the primary fields $\Phi_{m,n}$ and $V_{a_L}$ in the matter and Liouville sectors respectively
and $c$ ($\bar{c}$) is the $(-1)$-weight ghost field. The conformal dimension of 
$\Phi_{m,n} = \Phi_{q-m, p-n}$ ($1\le m < q, \; 1\le n < p$) 
is 
$ \Delta_{m,n} = \alpha_{m,n} (q_M + \alpha_{m,n})$, where
\be
\alpha_{m,n} = \frac{|m/\beta-n \beta|-q_M}{2} = \frac{|m p-nq|-(p-q)}{2 p \beta}
\ee
and the Liouville primary field $V_{a_L}$  
has $\Delta^L_{a_L} = a_L(Q_L-a_L)$. The total conformal dimension vanishes $\Delta(\WW_{m,n}) = \Delta(\widetilde{\WW}_{m,n})=0$, as a consequence of the BRST symmetry, leading to the constraint
\be
a_L =\alpha_{m, n}+\beta^{-1}\;.
\ee
 
The conformal property of the Liouville gravity induces the gravitation scaling property, defining the reaction of the theory on the re-scaling of the (bulk) cosmological constant $\mu$. 
The partition function $Z_L$ on the sphere has the following scaling behavior~\cite{KPZ}
\be
Z_L \sim \mu^{Q_L/b}\;,
\ee
so that its scaling dimension  (g-dim) is
$Q_L/b=(p+q)/q$, while the scaling dimensions of the physical fields~\eqref{phys-fields}
are equal to $ a_L/b$. One can assign  g-dim  to the coupling constant $\lambda_{mn}$ accompanying
fields in the MLG correlation numbers generating function 
$Z_L(\lambda)=\langle \exp \sum_{m,n} \lambda_{mn} \WW_{m,n} \rangle$, where $\lambda=\{\lambda_{mn}\}$: 
\be \label{eq:scaling}
[\lambda_{mn}]= \frac{p+q -|m p-nq|}{2 q}\;,
\ee
so that 
$ (p+q)/q =[\lambda_{mn}] + \alpha_{m,n}/\beta+\beta^{-2}$.

The scaling properties play a key role
in relating the observables of the matrix models with those of the Liouville gravity.
For example, in the spherical topology the Douglas equation~\eqref{Douglas-eqv} 
is conveniently formulated using the action principle~\cite{Ginsparg:1990zc}
$ \partial S(u) / \partial u_i= 0$
for the so-called Douglas action 
\be\label{eq:pre-S}
 S(u) =  \Res_{y = \infty} \left(Q^{\frac{p+q}{q}} + \sum_{m,n} t^{(m n)} Q^{\frac{|pm-qn|}{q}} \right)\;.
\ee 
Here the summation over $m$ and $n$ is restricted to the region 
$1\le m < q, \; 1\le n < p$ 
modulo equivalence $t^{(m n)} = t^{(q-m, p-n)} $. 
If one assigns g-dim $[Q]=1/2$, then taking into account that each term in \eqref{eq:pre-S} has the same g-dim 
one gets 
$[t^{(m n)}]=[\lambda_{m n}]$ defined in \eqref{eq:scaling}.
This allows to relate the coupling constants  $\lambda_{m n} $ of the Liouville gravity 
with the parameters $t^{(m n)} $ of the matrix models, up to some {\it resonance terms} to be discussed shortly.
 
 The susceptibility condition imposes
\be \label{eq:susceptibility}
\frac{\partial^2}{\partial x^2} \FF = v_1^*\;,
\ee
where $\FF$ is the free energy of the matrix model  
and  $v_1^*$  is a specific solution to the Douglas equation. 
Since
$\FF$ has the same g-dim as the Liouville partition function $Z_L$, 
$x$ becomes the parameter of the highest  g-dim which is equal to  $(p + q -1)/q$;
$[x]=[t^{(m_1,n_1)}]$ with $m_1,n_1$ subject to the constraint $|pm_1-qn_1|=1$.   

The free energy  $\FF(t)$ has contributions from all genus partition functions. To get a particular genus part one may 
introduce a formal genus parameter both in the free energy and in the Douglas string equation (leading to the genus expansion
of its solution)
and to combine then relevant terms.  
For the genus 0 case, according to this procedure, we should replace $\d$ in $Q$ in \eqref{eq:pre-S} by a commuting coordinate $y$.    
Using the notation $\theta_{i,k}$ introduced in \eqref{eq:theta_ik} one can write the action $ S(u)$
in the form 
\be\label{eq: S}
 S(v, t) =  \sum_{i=1}^{q-1}  t_i v_i  - H(v)\;, ~~~
 H(v, t) =  \theta_{p_0, s+1}   - \sum_{k \ne 0} t_{i,k}   \theta_{i,k}\,,
\ee 
where $p=sq+p_0$ with  a non-negative integer $s$. 
The terms  $\theta_{i ,k\ne 0}$ are collected in  $H(v)$. 
We note that the  indices $i,k$ and $m,n$ are related as
$i = |pm -qn| $ (mod $q$) and  $k = (|pm -qn| -i)/q \ne0$. 
In what follows, in spite of this one-to-one correspondence,  we keep sometimes  (for convenience) double labeling $ t_{i,k} ^{(m n)}  $ 
instead of $ t_{i,k}$.  We note that $[t_i^{(m_i  n_i)}] = [\lambda_{m_i  n_i}]$ 
and $ [t_{i,k} ^{(m n)}] = [\lambda_{m  n}]$.

Even though g-dim of $t^{(mn)}$ and $\lambda_{mn}$ are the same,
one cannot identify these two quantities. 
There are two reasons which prohibit such simple identification:
the different normalization of the two approaches 
and the appearance of the resonances between operators, 
emerging from contact term interactions and not violating the scaling property
\begin{equation}\label{resonance-transform}
t^{(m, n)} = \lambda_{m, n} + \sum A_{(m_1, n_1),(m_2, n_2)}^{(m, n)} \lambda_{m_1, n_1} 
\lambda_{m_2, n_2} + \cdots\;.
\end{equation} 
Here $A_{(m_1, n_1),(m_2, n_2)}^{(m, n)}$ are dimensionless constants and the sum goes over all possible
combinations respecting the scaling. Since powers of $\lambda_{11} \propto \mu$  provide scaling dimensions, it is convenient to reformulate the resonance transformation as follows
\begin{align}
t^{(mn)}\,=\,\,&\lambda_{mn}+ A_{mn} \mu^{\delta_{mn}} + \!\!\!\!\!\sum_{m_1,n_1}^{\delta_{m_1n_1}\leq
\delta_{mn}} \!\!\!\!A^{m_1n_1}_{mn}\mu^{\delta_{mn}-\delta_{m_1n_1}} \lambda_{m_1n_1}
+\nonumber\\
&+\!\!\!\!\sum_{m_1,n_1,m_2,n_2}^{\delta_{m_1n_1}+\delta_{m_2n_2}\leq
\delta_{mn}} \!\!\!\!\!\!A^{m_1n_1,m_2n_2}_{mn}\mu^{\delta_{mn}-
\delta_{m_1n_1}-\delta_{m_2n_2}} \lambda_{m_1n_1}\lambda_{m_2n_2}+\cdots \;,
\end{align}
where the non-vanishing coefficient appear when 
the power of $\mu$ is a non-negative integer. 

There are different solutions to the Douglas equation 
$\d S(v)/\d v_{i}|_{0^*} =0$,
where the condition ${0^*}$ stands for 
the case, where all the couplings $\lambda_{mn}$ vanish 
except $\lambda_{11}$. 
It is known~\cite{Belavin:2014hsa} that  the minimal gravity
is described by one particular  solution such that $v_{i>1}|_{0^*} =0$.  
In what follows, we call this {\it on-shell solution} 
and denote it $v^*$. Thus $v^*_{i>1} =0$ and the value $v_1^*$ is determined from the Douglas equation.

On the other hand, 
$   {\partial S}/{\partial v_1}=0$ is the integrated form of \eqref{Douglas-eqv}, 
whose integration variable is $x$.   
Since  $x$  and $t_1$ have  the same g-dim, 
two variables are closely related.
For later convenience, we define $x$
\be
\label{eq:x(v)}
x := \frac{\d H(v)}{\d v_1}\;,
\ee 
which gives implicit relation between $v_i$'s and $x$. 
If one takes the derivative with respect to $x$, 
eq.\eqref{eq:x(v)} reduces to eq.\eqref{Douglas-eqv}.
Since $\d S(v)/\d v_1= t_1 -x$,
we have $x={\d H(v)}/{\d v_1}|_{0^*}=t_1|_{0^*}$ 
as the one-shell solution.
The distinction of $x$ from $t_1$ is significant because 
the functional relation of $v$ with $x$ holds even off-shell,
where $t_1$ (considered as coupling constant) is to be identified with 
$\lambda_{m_1 n_1}$. 
 
The free energy $\FF$ has to satisfy the susceptibility condition 
\eqref{eq:susceptibility}. 
From this property, the free energy is~\cite{Belavin:2014cua}
\be 
\FF(t) = \frac{1}{2}\int_0^{v^*} c_{\beta\gamma}^{\alpha} \frac{\d S}{\d v_{\beta}}
 \frac{\d S}{\d v_{\gamma}} \dd v_{\alpha}\;.
\ee
Because the integrand is the closed one-form~\cite{Belavin:2013nba}, taking into account the property of the relevant solution $v^*$,
it is convenient to chooses the integration path  $d v_{i>0} =0$,
then the free energy has simpler form
\be 
\label{eq:Bulk-sphere}
\FF(t) = \frac{1}{2}\int_0^{v_1^*} c_{\beta\gamma}^1 \frac{\d S}{\d v_\beta}
 \frac{\d S}{\d v_{\gamma}} \dd v_1\;.
\ee
Bulk correlation numbers are obtained 
by differentiating with respect to the corresponding coupling constants.
Thus, \eqref{eq:Bulk-sphere} is the generating function of the bulk correlation on the sphere.

\subsection{Bulk generating function  on the disk} 
\label{sec:2_disk} 

The  bulk free energy on the disk is obtained 
using the free energy idea of one matrix model.
It is shown in \cite{Ishiki:2010wb}
that the boundary free energy with $h \ge 1$ holes (boundaries) is given by 
\begin{align}
\FF_h(t)=\frac{1}{h!} \langle (-{\rm Tr}\log C(M))^h  
\rangle_c
=\frac{1}{h!} \left\langle \left( \int_0^\infty \frac{dl}{l}~
{\rm Tr}\; e^{-lC(M)}\right)^h  
\right\rangle_{\!\!\!c}\;,
\label{eq:boundary-free-energy-h-holes}
\end{align}
where the expectation value is evaluated 
with respect to the bulk interaction 
and the subscript $c$ refers to the connected part.  
Tr acts on the matrix components. 
With the one boundary loop (h=1)
one has the free energy  on the disk.
\be 
\label{eq:F1}
\FF_1(t)= -\langle {\rm Tr} \log(\mu_B-M) \rangle_c
= \int_0^{\infty} \frac{d l}{l}   e^{-\ell  \mu_B}  ~
 {\rm Tr}\; \left \langle e^{lM} \right\rangle\;.
\ee
Here we use the prescription  $C(M)=\mu_B-M$ 
which describes the case with no boundary operators
with  $\mu_B $   boundary cosmological constant.
The derivative of  $\FF_1$ with respect to $\mu_B$ 
is  the one-point resolvent expectation value. 
It is noted that the integration over $l$ is the Laplace transform, 
and the integration path is chosen form 0 to $\infty$
assuming the contribution due to 
$ {\rm Tr}\; \left \langle e^{lM} \right\rangle$ is convergent.
If not, the integration range is properly re-defined so that the 
integration converged. 

The expectation value  
${\rm Tr}\; \left \langle e^{lM} \right\rangle $ is 
the main element for finding the  free energy on the disk.
At the continuum limit, $M$ is replaced by $Q$.
In \cite{MSS} the element is treated as the expectation value of the 
loop operator $\left \langle w(l) \right\rangle $.
We will  denote the expectation value as  ${\cal W} (l) $.
\be 
\label{eq:W}
{\cal W}(l) = \int\limits_{x_1}^{\infty} \dd x \, 
\langle x| e^{ l Q} |x\rangle 
= \int\limits_{x_1}^{\infty} \dd x \int_{i\RR} \dd y~ e^{l Q(y, v(x))}\;,
\ee
where $Q$ is given in $x$-representation. 
Tr becomes the integration over $x$ and
the integration interval  $x_1$
is to be identified with $t^{(m_1,n_1)}$.

The generating function on the disk with genus 0
is obtained if $\d$ is replaced by $y$ in  $Q$.
Since $y$ represents the  Fourier space,
we have put $Q =Q(y, v(x))$ and integration  of $y$ is performed over $i \RR$.
In addition, the flat coordinates $v$ have  $x$-dependence 
through  \eqref{eq:x(v)},
which is different from on-shell value $v^*$. 
On the sphere the bulk generating function 
does not contain explicit $x$-dependence of $v$.
However, on the disk  explicit $x$-dependence is necessary.
On the disk  we need $Q(y, v(x, t))$  and more information about $x$ and coupling constant dependence is required in order to get the generating function $ {\cal W}(l,t)$.

The bulk correlation on the disk is given as 
a certain derivatives of the bulk generating function  \eqref{eq:W}.
For example, one-point correlation is given as 
\be  
\label{eq:one-pt-disk}
\langle O_{m, n} \rangle_D = \left. \pd{\lambda_{m, n}} 
{\cal W}(l) \right|_{0*} \;,
\ee
where $0^*$ has the same meaning as in the spherical case.
The result~\eqref{eq:one-pt-disk} 
is to be compared with  FZZ result in  \cite{Fateev:2000ik}
\begin{equation}\label{FZZ}
W_{\alpha_L}(l_0) = \frac{2}{b} \left( \pi \mu \gamma(b^2) \right)^{(Q_L-2\alpha)/2b}
 \frac{\Gamma(2\alpha  b - b^2)}{\Gamma(1+1/b^2 - 2\alpha/b)} K_{(Q_L-2\alpha_L)/b} (\kappa l_0)\;,
\end{equation}
where   $\kappa^2 = \mu/\sin (\pi b^2)$
and the order of the Bessel function is given as 
 $ \nu_{m, n} = \frac{ Q_L -2\alpha_{m, n}}b   = \frac{|mp - nq|}{q}$.
 More specifically for the minimal model, FZZ result has the form
 \begin{equation}\label{eq:FZZ}
W_{\alpha_L}(l_0)  = 
e_{\alpha_L} \kappa^{\nu_{m, n} }K_{\nu_{m, n} } (\kappa l_0)\;,
\end{equation} 
where $e_{\alpha_L}$ is a certain 
numerical constant  independent of $\kappa$ and $l_0$.\footnote{We note that the length parameters $l$ and $l_0$ in \eqref{eq:W} and \eqref{FZZ} are different. The relation between them is model dependent and will be fixed in our concrete examples. This is because there is
 no canonical normalization} Explicit checks aimed at testing  \eqref{eq:one-pt-disk} against \eqref{eq:FZZ} will be the subject of the next two sections. 
%

\section{Unitary models}
\label{sec:3}

In this section we compute one-point functions
for the unitary series of Minimal Gravity in the
``dual'' aproach using 
the existing results on the resonance transformation.
We compare our results with the ones from  
the Liouville gravity.

\subsection{M(3/4): Ising model -- the simplest unitary case}
\label{sec:3_M(3/4)}

According to the general framework discussed above,
the Ising model is governed by the $A_2$ Frobenius manifold with the Lax polynomial
 $Q= y^3 + v_1 y +v_2$, where $v_1$ and $v_2$ are the flat coordinates. 
The structure constants are $c^{122}=1, c^{111}= -v_1/3$ and the nontrivial component
of the metric is $\eta_{12}=1$.
The Douglas action has the form  
\begin{align}
S= t_1^{(11)} v_1 +  t_2^{(12)} v_2- H \;,~~~~
H=\theta_{1, 2} - t_{2,1}^{(13)} \theta_{21}\;,\\
\theta_{1, 2}= (-v_1^4 + 18 v_1 v_2^2)/36\;,~~~
\theta_{2,1} = (-v_1^3 + 9 v_2^2)/18\,,
\end{align} 
where $(m,n)$ index has the $Z_2$ symmetry: 
$t_1^{(11)} = t_1^{(23)}, ~t_2^{(12)}=t_2^{(22)}, ~t_{2,1}^{(13)}=t_{2,1}^{(21)} $.
The gravitational dimensions of the coupling constants are
\be
[t_1^{(11)}]=1\;,\;\; [t_2^{(12)}]=5/6\;, \;\;[t_{2,1}^{(13)} ]=2/6\;.
\ee
Since $t_1^{(11)}$ has g-dim 1 the following  resonance terms are allowed
\begin{align}
\label{eq:34-resonace}
&t_1^{(11)} =\lambda_{11} + A^{(11)}_{(21),(21),(21)} \left( \lambda_{13}  \right)^3\;, \\
&t_2^{(12)}= \lambda_{12}\;, \\
&t_{2,1}^{(13)} = \lambda_{13}\;.
\end{align}
The derivatives of  $S$ are 
\begin{align}
\label{eq:S1}
& \frac{\partial S}{\partial v_1} = t^{(11)}  -x\;, ~~~\text{where}~~~
x=- \frac 1{9   }  v_1^3 
+ \frac 12 v_2^2  +\frac16   t^{(13)}_{21}  { v_1^2} \;,
\\
\label{eq:S2}
& \frac{\partial S}{\partial v_2} =   t_2^{(12)}  
+  t^{(13)}_{21}  v_2 - v_1v_2\;.
\end{align}   
The Douglas equation reads 
\be
\frac{\d S}{\d v_1} = \frac{\d S}{\d v_2} = 0\;.
\ee 
From~\eqref{eq:S1},~\eqref{eq:S2} it follows that the on-shell solution (all $t = 0$ except $t^{(11)} \sim \mu$)  
is $v_2^*=0, \; (v_1^*)^3 = 9 \lambda_{11}$.
In addition, $x$-dependence of $v_i$'s is fixed by the 
second equation in~\eqref{eq:S1}.

Bulk generating function on the sphere is \cite{Belavin:2014cua}
\be 
\label{eq:A2-sphere}
\FF(t) = \frac12 \int^{v_1^*} dv_1 \left( 
\left(\frac{\partial S}{\partial v_1}\right)^2 
- \frac{v_1}3  \left(\frac{\partial S}{\partial v_2}\right)^2
 \right)\;.
\ee
One can check that the bulk one-point correlation numbers on the sphere vanish. 
Explicit evaluation shows that  
\be 
\label{eq:12}
   \langle O_{12} \rangle 
   = \left. \frac{\partial \FF(t) }{\partial \lambda_{12} } \right|_{0^*} 
=  \int_0^{v_1^*} dv_1 \left( - \frac{v_1}3  \right)  (-v_1v_2^*) \;,
\ee
which is identically zero  
due to the on-shell value $v_2^* =0$ 
deternined in   \eqref{eq:S2}. 
Note that $ [O_{12}]=3/2$ and $  \langle O_{12} \rangle $ is non-analytic in $\mu$ and
this correlator
must therefore vanish.  
On the other hand, $[ O_{13}]= 2$ and   
\be
\label{eq:13}
 \langle O_{13} \rangle  
 = \left. \frac{\partial \FF(t) } {\partial \lambda_{13}} \right|_{0^*} 
=  \int_0^{v_1^*} dv_1   
\left(  \mu + \frac{v_1^3}9 \right)  \left( -   \frac{v_1^2} 6  \right) 
\propto \mu^2\;.
\ee
The non-vanishing contribution is analytic in $\mu $ 
and is not universal, so that it can be discarded. 
This shows that $\langle O_{13} \rangle  $= 0 (mod  $\mu ^2$).
One may also check that two-point correlator 
$ \langle O_{22} O_{21}\rangle $ 
vanishes automatically since $v_2^*=0$.

Bulk one-point correlation on the disk is defined according to \eqref{eq:one-pt-disk}.
Its generating function  is 
$ {\cal W}(l, t) = \int\limits_{\mu}^{\infty} \dd x \, 
\langle x| e^{ l Q} |x\rangle  
$. The polynomial $ Q(y, v_1) = y^3 + v_1 y$ can be written in terms of Chebyshev polynomial
$T_3 (x) = 4 x^3 -3x$
\be
Q(y, v_1) = 2 i (v_1/3)^{3/2} T_3 \left(\frac{i y \sqrt{3}}{2 \sqrt{v_1}}\right)\;.
\ee
Therefore, we have 
\be 
\label{eq:34D-11}
\langle O_{1 1} \rangle_D 
 =
\int_{\RR} \dd y \; e^{il(y^3 + v_{1*} y)} = \frac{2 v_{1*}^{1/2}}{3} K_{1/3}(2l (v_{1*}/3)^{3/2})\;,
\ee
where  on-shell value  $v_{2*}=0, v_{1*} = (9\mu)^{1/3}$ is used.

To evaluate other one-point functions we need to compute
 $\frac{\d Q(y, v(x))}{\d \lambda_{m, n}}$ on-shell.
For this purpose we use the Douglas equation \eqref{eq:S1} and \eqref{eq:S2}
\be
 \left.  \frac{\d v_1}  {\d t_{2,1}^{(21)} }\right|_{0^*}=  -\frac{1}{2} \;,~~~
 \left. \frac{\d v_2}{\d t_2^{(22)} }\right|_{0^*}= \frac {1}{v_1}\;.
\ee
This shows that 
\be 
\langle O_{2,2} \rangle_D 
  =  l 
\int\limits_{ \mu}^{\infty}  \frac{\dd x}{v_1} ~ 
 \int_{\RR} \dd y e^{i l  (y^3 + v_1 y)}   
 =
 -\frac{4lv_{1*}}{3 z} K_{2/3}(z\mu^{1/2})\;, \; z = 2 l \sqrt{3}\;,
\ee
where we used the integration formulae from the appendix~\ref{sec:app}.
This coincides with the FZZ result~\eqref{FZZ}.
Similarly we have  
\be
\langle O_{1,3} \rangle_D  = -\frac{4}{9 z} i l v_1^{5/2} K_{5/3}(z\mu^{1/2})\;.
\ee
In particular we find the relation between $l$ and $l_0$ from~\eqref{eq:FZZ}:
$l/l_0 = 2^{3/4}/\sqrt{3}$.

\subsection{M(q/q+1): towards the general case}
Now we describe the general unitary minimal model
$(q,p) = (q, q+1)$.
Corresponding Frobenius manifold is $A_{q-1}$ and the Lax operator is
 $Q=y^q + \sum_{i=1}^{q-1} u_i y^{q-1-i}$.
The flat coordinates of the  $q-1$-dimensional  Frobenius manifold  
are given as $v_i = \theta_{i,0} =u_i + \cdots$ and the metric is  
$\eta_{ij}=\delta_{i+j, q}$.

In general the structure constants are complicated.
However, the bulk correlation generating function 
needs the structure constants $c^1_{ij}$ on-shell only, which follows
from the condition $v_{i>1}^*=0$.
It was shown in~\cite{Belavin:2014cua} that $c^1_{ij}$ on-shell has non-vanishing components 
when $i=j$ only and is given explicitly as  
 \be
 \label{eq:c1ij-on-shell}
 c^1_{ij} =\delta_{ij} ~c_i~ v_1^{i-1}\;,
 \ee
 $c_i$ being  a constant. 
   
We have the action $S(v)$ of the form 
\be
S(v) = \sum _i v_i t_i^{(i,i)}  -H(v) \;, ~~~~
H(v) = \theta_{1,2}(v)  -\sum_{j =1}^{q-1} \sum_{k\ge 1} 
t_{j,k}^{(m,n)} \theta_{j,k}(v) \;,
\ee
where $(k = m-n, j=m) $ or $(m=j, n=j-k)$. The parameter $t$ 
has the symmetry 
 $t_{i, k}^{(m,n)} = t_{i,k} ^{(q-m, p-m)}$.  
 The highest g-dim is 1 and therefore we have $[x] =[\lambda_{11}]$. 
The derivative of the action is
$ {\partial S}/{\partial v_i} = t_i^{(i,i)} -   {\partial H(v)}/{\partial v_i}$ 
with 
\be
\frac{\partial H (v)}{\partial v_i}= \frac{\partial \theta_{1,2}(v) }{\partial v_i}
-\sum_{j=1}^{q-1} \sum_{k\ge 1} 
t_{j,k}^{(m,n)}  \frac{\partial  \theta_{j,k}(v) }{\partial v_i}\;.
\ee
The Douglas equation $ {\partial S}/{\partial v_{i}} = 0$ has the on-shell solution 
$v^*_{i >1}=0$ and on-shell value of $Q$ is given 
in terms Chebyshev polynomial $T_q$ ~\cite{DiFrancesco1991}
\be
 Q(y, v_1) = 2 (-i)^q \left( \frac{v_{1}^*}{q}\right)^{q/2} T_q \left(i y \frac{q^{1/2}}{2 v_{1}^{*1/2}}\right) = y^n + y^{n-2} v_{1}^* + \cdots.
 \ee 
The derivative of the action for $(v_{i>0} = 0)$ can be found explicitly using
\bal \label{eq:theata-derivative}
&\frac{\partial  \theta_{j,k}(v) }{\partial v_i} =
  \delta_{j,i} x_{j,k}  (-v_1/q)^{kq/2} ~~{\rm when}~~k~{\rm even}\;,\\  
&\frac{\partial  \theta_{j,k}(v) }{\partial v_i} = \delta_{j,q-i} y_{j,k}
 (-v_1/q)^{(k-1)q/2+j } ~~{\rm when}~~k~{\rm odd}\;,
\eal
 where 
\bal
x_{j,k}= \frac{1}{\left(\frac{j}{m}\right)_{\frac{k}{2}} \left(\frac{k}{2}\right)!}\;, \qquad
y_{j,k}= -\frac{1}{\left(\frac{j}{m}\right)_{\frac{k-1}{2}} \left(\frac{k-1}{2}\right)!} \;\\ 
\eal 
and $(a)_n = \Gamma(a+n)/\Gamma(a)$ is a Pochhammer symbol.
The bulk generating function on the sphere  is 
\be
\FF(t) =  \sum_{i=1}^{q-1}  \frac{ c_i}2  \int^{v_1^*}_0 ~
dv_1 ~  v_1^{i-1} 
\left(\frac{\partial S}{\partial v_i}\right)^2 \;, 
\ee
where the integration path is along $v_{i>0} = 0$. 
In addition,  the structure constant $c^1_{jm}$ in~\eqref{eq:c1ij-on-shell} is used.
 Differentiating this function and requiring one-point function vanishing,
 diagonality of two-point function and other fusion rules, the 
 resonance transformation were found in~\cite{Belavin:2014cua}.
To this end,  we introduce new integration variable $w$ 
\be
w=2\left(\frac{v_1}{v_{1}^*}\right)^q-1\;.
\ee
The integration now looks like
\be
  \int^{v_1^*}_0 ~\frac{dv_1}{v_1} \left(\frac{v_1}{v_{1}^*}\right)^q 
  \to \frac1{2q}\int_{-1}^1dw\;.
\ee
In the new variable the fusion rules become the orthogonality conditions for Jacobi polynomials
\be
\frac1q\int_{-1}^1 {dw}(1+w)^\beta P_m^{(0, \beta)} P_n^{(0, \beta)}  
= \delta_{mn} \frac{2^{\beta+1}} {2n  +\beta+1} \;.
\ee
In particular, if $\beta=0$  Jacobi polynomial reduces to Legendre polynomial. \\

The generating function on the disk is given in \eqref{eq:W}
and its one-point correlation is given in \eqref{eq:one-pt-disk}. Explicitly one gets
(if $(m,n) \ne (1,1)$)~\footnote{In the exponent the sign should be picked in a way that the integral converges.
That is $- l Q(y, v)$ if $q$ is even and  $i l Q(y,l)$ if $q$ is odd.}
\be \label{eq:unit-one-pt}
\langle O_{m,n} \rangle = l \int_{\mu}^{\infty} \dd x \int_{\RR} \dd y \; \frac{\d Q(y, v)}
{\d \lambda_{m,n}} \; e^{l Q(y, v)}\;.
\ee 
Pre-exponent is evaluated using
\be
\frac{\d Q(y)}{\d v_{\alpha}}\frac{\d v_{\alpha}}{\d \lambda_{m,n}} = 
\frac{\d Q(y)}{\d u_a}\frac{\d u_a}{\d v_{\alpha}} \frac{\d v_{\alpha}}{\d \lambda_{m,n}}\;.
\ee
We use the expressions for the second and third terms from the paper~\cite{Belavin:2014cua}
\bal
&\d v_{\alpha}/\d \lambda_{m,n} = -\d_{v_{\beta}} S^{(m, n)} \left[\left(\frac{\d^2 S_0}{\d v^2}\right)^{-1}\right]^{\alpha, \beta}\;,\\
&\left[\left(\frac{\d^2 S_0}{\d v^2}\right)^{-1}\right]^{\alpha, \beta} = \frac{1}{v_1}
\left(\frac{-v_1}{q}\right)^{1-q+\alpha} \delta^{\alpha\beta}\;. \\
\eal
Here $S^{(m,n)} :=\d S / \d t_{i, j}^{(m,n)}$ and their derivatives
 $\d S^{(m,n)}/\d v_a$ are given in terms of the Jacobi polynomials~\cite{Belavin:2014cua}
\bal \label{eq:dSdv}
&\left.
 \frac {\d S^{(m,n)}} {\d v_a }  
\right|_{0^*} 
= -\delta_{a, m}\left(\frac{m}{q}\right)_{m-n}^{-1} \left(\frac{-v_{1}^*}{q}\right)^{\frac{m
-n}{2}q} P_{\frac{m-n}{2}}^{(0, \frac{m}{q}-1)} (w)\;, \\
&\left. 
 \frac {\d S^{(m,n)}} {\d v_a } 
\right|_{0^*} 
= \delta_{q-a, m}\left(\frac{m}{q}\right)_{m-n}^{-1} \left(\frac{-v_{1}^*}{q}\right)^{\frac{m-
n-1}{2}q}\left(\frac{-v_{1}}{q}\right)^{m} P_{\frac{m-n-1}{2}}^{(0, \frac{m}{q})} (w) \;.
\eal
From these expressions it is clear that differentiating wrt $\lambda_{mn}, \; m>n$ changes only
$v_{m}$ if $m-n$ is even and $v_{q-m}$ otherwise. Using the formulae for $\d u_a/ \d v_b$
from~\cite{Belavin:2014cua} we evaluate
\be
\frac{\d Q(y, v)}{\d v_{q-\gamma}} = (-i)^{\gamma-1} \left(\frac{v_1}{q}\right)^{\frac{\gamma-1}{2}}
 T^{\prime}_{\gamma}\left(iy \frac{\sqrt{q}}{2\sqrt{v_1}}\right)\;.
\ee
Thus, using expression~\eqref{eq:unit-one-pt} and formulae above we obtain
\bal \label{oneptfn2}
\langle O_{m,n} \rangle &\sim   l \int_{\mu}\dd x\int_{\RR}\dd \tilde{y} \; v_1^{\frac{m-q}{2}}
\, v_{1}^{*\frac{m-n}{2}q} \, P^{(0, \frac{m}{q}-1)}_{\frac{m-n}{2}}
(w) \, T^{\prime}_{q-m}(\tilde{y}) \,
\exp\left[-z T_q(\tilde{y})\right], \; (m-n)\text{ even}\;,\\
\langle O_{m,n} \rangle &\sim   l \int_{\mu}\dd x\int_{\RR}\dd \tilde{y} \; v_1^{\frac{m}{2}}
 \,v_{1}^{*\frac{m-n-1}{2}q}  \, P^{(0, \frac{m}{q})}_{\frac{m-n-1}{2}}
(w) \, T^{\prime}_{m}(\tilde{y}) \,
\exp\left[-z T_q(\tilde{y})\right], \; (m-n)\text{ odd}\;,
\eal
where $z = 2 (v_1/q)^{q/2} l$ and $\tilde{y} = \sh s = y \sqrt{q}/(2\sqrt{v_1})$. Using formulae 
from the appendix~\ref{sec:app} if $q$ is even  we get
\be \label{eq:int-even}
\int_{\RR} \dd y \, \d Q(y,v)/\d \lambda_{m,n} \, e^{-l Q(y,v)} = \\
\ee
\label{my-label}
\begin{center}
\begin{tabular}{l | c | r}
 & m-n \text{ even} & m-n \text{ odd}   \\ \hline
m \text{ even} & 0    & 0      \\ \hline
m \text{ odd} & $-C_1[v_1, v_{1}^*] K_{1-\frac{m}{q}}(z)$  & $C_2[v_1, v_{1}^*] K_{\frac{m}{q}}(z)$
\end{tabular}
\end{center}
where
\bal
C_1[v_1, v_{1}^*] = 4\frac{q-m}{q^2} \left( \frac{m}{q} \right)^{-1}_{m-n}
 \left(\frac{v_1}{q}\right)^{\frac{m-q}{2}}
 \left(\frac{v_{1}^*}{q}\right)^{\frac{m-n}{2}q} P^{(0, \frac{m}{q}-1)}_{\frac{m-n}{2}}(w) = \\
const \, v_1^{\frac{(m-n-1)q + m}{2}} + const \, v_1^{\frac{(m-n-3)q + m}{2}} v_{1}^{*q} + \cdots \\
C_2[v_1, v_{1}^*] = 4\frac{m}{q^2} \left( \frac{m}{q} \right)^{-1}_{m-n}
 \left(\frac{v_1}{q}\right)^{\frac{m}{2}}
 \left(\frac{v_{1}^*}{q}\right)^{\frac{m-n-1}{2}q} P^{(0, \frac{m}{q})}_{\frac{m-n-1}{2}}(w) = \\
const \, v_1^{\frac{(m-n-1)q + m}{2}} + const \, v_1^{\frac{(m-n-3)q + m}{2}} v_{1}^{*q} + \cdots \\
\eal
and if $q$ is odd
\be \label{eq:int-odd}
\int_{\RR} \dd y \, \d Q(y,v)/\d \lambda_{m,n} \, e^{i l Q(y,v)} = \\
\ee
\label{my-label}
\begin{center}
\begin{tabular}{l | c | r}
 & m-n \text{ even} & m-n \text{ odd}   \\ \hline
m \text{ even} & $-(-1)^{\frac{m-n}{2}} \, \cos(\frac{\pi}{2q}) \, C_1[v_1, v_{1}^*] \, K_{1-\frac{m}{q}}(z)$
    & $-i(-1)^{\frac{m-n-1}{2}} \, \sin(\frac{\pi}{2q}) \, C_2[v_1, v_{1}^*] \, K_{\frac{m}{q}}(z) $     \\ \hline
m \text{ odd} & $i(-1)^{\frac{m-n}{2}} \, \sin(\frac{\pi}{2q}) \, C_1[v_1, v_{1}^*] \, K_{1-\frac{m}{q}}(z) $  & $-(-1)^{\frac{m-n-1}{2}} \, \cos(\frac{\pi}{2q}) \, C_2[v_1, v_{1}^*] \, K_{\frac{m}{q}}(z)$
\end{tabular}
\end{center}

To get one-point functions we need to integrate $\int_{\mu} \dd x$ the results
 of~\eqref{eq:int-even} and~\eqref{eq:int-odd}. At the moment we are not aware how to perform it 
in the general case, however in all concrete examples we get
\bal
\langle O_{m,n} \rangle &\sim \mu^{\nu/2} \, K_{\nu}(l \mu^{1/2}), \; \nu = m-n + \frac{m}{q}\;, \\
\langle O_{m,n} \rangle &\sim 0, \; m, q \in 2\mathbb{Z}\;.
\eal
At first sight
it appears that the second case is in contradiction with the FZZ result.
One of the reasons may be the symmetric minimal model boundary conditions chosen by matrix model
integrals, however to check this we need to fix exact normalizations between MLG and Douglas
equation approach. This phenomenon resembles also certain nonvanishing correlators appearing in the spherical case,
which should vanish in the CFT approach (for the discussion see e.g.~\cite{Belavin:2013nba, Belavin:2014cua}).

For $m=n$ and $m=n+1$, Jacobi polynomial reduces to constant and the integrals can be taken
for all q and m 
\be
C_1[v_1, v_{1}^*] = 4\frac{q-m}{q^2} 
 \left(\frac{v_1}{q}\right)^{\frac{m-q}{2}} , \;\;\;\;\; C_2[v_1, v_{1}^*] = 4 \frac{m}{q^2}
\left(\frac{v_1}{q}\right)^{\frac{m}{2}}\;.
\ee
Therefore we get if q is even, then m is odd and
\be \label{eq:int-even1}
\langle O_{m,n} \rangle = \\
\ee
\label{my-label}
\begin{center}
\begin{tabular}{l | c | r}
 & m=n  & m=n+1   \\ \hline
m \text{ odd} & $-4\frac{q-m}{\tilde{z} q^2} \left(\frac{v_{1}^*}{q}\right)^{\frac{m}{2}} K_{\frac{m}{q}}(z_*)$ 
 & $4 \frac{m}{\tilde{z} q^2} \left(\frac{v_{1}^*}{q}\right)^{\frac{m+q}{2}} K_{1+\frac{m}{q}}(z_*)$
\end{tabular}
\end{center}
where $z_* = z|_{v_1 = v_{1}^*}$ and $\tilde{z} =  z_*/(2l\sqrt{\mu}) = \sqrt{\frac{q}{(q+1)(2q+1)}}$.
Similarly for odd $q$ we obtain

\be \label{eq:int-odd1}
\langle O_{m,n} \rangle = \\
\ee
\label{my-label}
\begin{center}
\begin{tabular}{l | c | r}
 & m=n & m=n+1   \\ \hline
m \text{ even} & $-4\cos(\frac{\pi}{2q})\frac{q-m}{\tilde{z}_* q^2} \left(\frac{v_{1}^*}{q}\right)^{\frac{m}{2}} K_{\frac{m}{q}}(z_*)$
    & $-4i \sin(\frac{\pi}{2q})\frac{m}{\tilde{z}_* q^2} \left(\frac{v_{1}^*}{q}\right)^{\frac{m+q}{2}} K_{1+\frac{m}{q}}(z_*)$     \\ \hline
m \text{ odd} & $4i\sin(\frac{\pi}{2q})\frac{q-m}{\tilde{z}_* q^2} \left(\frac{v_1}{q}\right)^{\frac{m}{2}} K_{\frac{m}{q}}(z_*)$  & $-4\cos(\frac{\pi}{2q}) \frac{m}{\tilde{z}_* q^2} \left(\frac{v_{1}^*}{q}\right)^{\frac{m+q}{2}} K_{1+\frac{m}{q}}(z_*)$
\end{tabular}
\end{center}
We also write down the partition function
\be
Z_B = 2 c \, \frac{v_{1}^{*\frac{q+1}{2}}}{l \tilde{z}_* q^{3/2}} \, K_{1+\frac1q}(z_*)\;,
\ee
where $c = 1$ for even q and $c = \cos(\pi/2q)$ for odd q.

\section{Non-unitary models}
 \label{sec:4}
 \subsection{M(2/2s+1): Lee-Yang series}
 \label{sec:4-lee-ynag}

We revisit first the Lee-Yang series on the disk in~\cite{Belavin:2010ba} 
and discuss then the normalization effect. 
The Lee-Yang series $M (2/ p)$ with $p=2s+1$ 
is described by one-matrix model and 
is based on 
$A_1$ Frobenius manifold with $Q = y^2 + v$.
The metric is trivially given by $\eta_{1,1}=1$. 
Nevertheless, the Douglas action is not trivial
\be
S(v) =   t_1^{(1,s)}   v  - H(v) \;, ~~~~
H(v) =\frac1{s+2} v^{s+2}  
 - \sum_{k=1}^{s-1}    \frac{t_{1,k}^{(1,n_k)}}{k+1} v^{k+1}\;,
\ee
where $n_k = s-k \ge 1 $. 
We use the fact that $  \theta_{1,k}(v) \sim v^{k+1}$
and put some normalization constant for later convenience.
The parameter $t$ has ${\mathbb Z}_2$ symmetry
$t_{1,k}^{(1,n_k)} = t_{1,k} ^{(1, p-n_k)}$ and  
g-dim of $t$ is $[t ^{(1,n )}]= (1+n )/2$. 
The highest g-dim is $(1+ s)/2$ and  therefore, 
$[t_1^{(1,s)}]=[x] =(1+ s)/2$, 
whereas $[t_{1, s-1}^{(1,1)}]=[\lambda_{11}]=1$. We remind that
 $\lambda_{1,1} =\mu$. 
Considering g-dim, the resonance transformation is as follows.
When $n $ is odd ($\ne 1$) and $[t ^{(1,n )}]$ is integer
\begin{align}
t^{(1,n)}
&= \lambda_{1,n}+ A_{(1,n)}^{(1,1)^{(n+1)/2}} \mu^{(n+1)/2}  
+  \sum_ {a=3, 5, 7 \cdots} 
A_{(1,n)}^{(1,1)^{(n-a)/2},(1,a) } \mu^{(n-a)/2}  \lambda_{1, a}
\nonumber\\
&~~~~~~~~
+ \sum_ {a+b=3, 5, 7 \cdots} 
A_{(1,n)}^{(1,1)^{(n-a-b-1)/2},(1,a) (1, b) } \mu^{(n-a-b-1)/2}  \lambda_{1, a}\lambda_{1, b}+ O(\lambda^3) \;.
\end{align}
When $n $ is even and $[t ^{(1,n)}]$ is half odd integer
\begin{align}
t_{1,k}^{(1,n)}
&= \lambda_{1,n}
+  \sum_ {a=3, 5, 7 \cdots} 
A_{(1,n)}^{(1,1)^{n/2-a}, (1,a) }\mu^{n /2-a}  \lambda_{1, 2a}
\nonumber\\
&~~~~~~~
+ \sum_ {a+b=odd} 
A_{(1,n)}^{ (1,1)^ { (n-a-b-1)/2},(1,a) (1, b) } \lambda_{1,1}^{ (n -a-b-1)/2}  \lambda_{1, a}\lambda_{1, b}+ O(\lambda^3)\;.
\end{align} 

The derivative of the action is
$ {\partial S}/{\partial v} \equiv t^{(1,s)} -   x $, 
where 
\be
x:=\frac{\partial H (v)}{\partial v}= v^{s+1} 
- t  ^{(1,1)} v^{s-1} -\sum_{k=1}^{s-2} t ^{(1,s-k)} v^k  \;,
\ee 
which presents the implicit dependence of  $v$  on $x$. 
Using the resonance transformations 
$x$ was computed in ~\cite{Belavin:2008kv} 
(including the resonance terms of $t^{(1s)}$)
\begin{align}
\label{eq:x-resonance}
\hat x(v)
&=x(v) + \lambda_{1s} -t^{(1s)} 
\nn\\
&=
\left(\frac{v_0^{s+1}}{(s+1) N_s}\right)  \left( \frac{P_{s+1}(z) -P_{s-1}(z)}{2s+1} \right)
-\lambda_{1n}\left( \frac{v_0^{s-n}}{N_{s-n}}\right) P_{s-n}(z) + O(\lambda^2) \;,
\end{align}  
where $P_n(z)$ are Legendre polynomials
with   $ z = v/v_0$ and $v_0 = \sqrt{\lambda_{(11)} \frac{2(2s-1)} {s(s+1)}}$. Here
$N_k =2^k \Gamma(k+1/2)/(\Gamma(k+1)\Gamma(1/2))$ is a nomalization factor
so that $ \frac{v_0^{k}}{N_{k}} P_{k}(z) = v^{k} + \cdots$ is a monic polynomial.
   
The bulk generating function on the sphere is  defined as 
\be
\FF(t) = \frac12  \int^{v _0}_0 ~dv  
~ \left(\frac{\partial S}{\partial v}\right)^2 .
\ee
Note that $v^*=\pm v_0$ since  $ x(\pm v_0) |_0 =0$. 
Here we choose $v^*=v_0$. 
The one-point function  $\langle O_{1,n} \rangle $ (for $2\le n \le s-1$)
is given as 
\be 
\frac{\d \FF(t)}{\d \lambda_{1,n} } 
\sim   \int^{1}_0 ~dz  ~    P_{s-n}(z) ~
 \left( \frac{ P_{s+1}(z)-P_{s-1}(z)}{2s+1} 
\right)\;,
\ee 
which vanishes due to the orthogonality of the Legendre polynomials.  
In the same way, two-point correlation satisfies the orthogonality.

On the disk, the generating function ${\cal W}(l)$ in  \eqref{eq:W} is used.
However, we need a systematic way to find the perturbative contribution 
using the expression  \eqref{eq:x-resonance}.
For this purpose, we interpret ${\cal W}(l)$ as
\be
{\cal W}(l) = \int_{\lambda_{1s}} dx_0 
\int_{i \RR}  dy    e^{  l (y^2 + v)} \;,
\ee
where we use the notation $x_0 = \hat x(v)|_{\lambda_{1n}=0}$ in  \eqref{eq:x-resonance}
\be
x_0 =\left(\frac{v_0^{s+1}}{(s+1) N_s}\right)  \left( \frac{P_{s+1}(z) -P_{s-1}(z)}{2s+1} \right).
\ee
Then, the perturbative contribution 
is due to $\delta \lambda_{1n}$ in  \eqref{eq:x-resonance}
and its effect on $\delta v$ is calculated  
by constraining $\delta \hat x =0$.

The correlation $\langle O_{1,s} \rangle_D  $ is simply given as 
\be 
\langle O_{1,s} \rangle_D  
=-    
\int_{i \RR}  dy    e^{  l (y^2 + v^*)} 
=-    \sqrt{ \frac{\pi}{l}} e^{  -l  v_0 } 
\propto \kappa^{ 1/2} K_{ 1/2}(\kappa l)\;,
\ee
where $  \kappa = v_0 $. We use $v^* =-v_0$ to make the integral  convergent.

Other correlations $\langle O_{1,n\ne1} \rangle_D$ 
are given as 
\be 
\langle O_{1,n} \rangle_D  
= 
\sqrt{  {\pi}{l}}
\int_0 \dd x_0 
\left(  \frac{\delta v}{ \delta  \lambda_{1,n}}  \right)_{\!0}~  e^{  l v}\;. 
\ee
The variation of $v$ is obtained from \eqref{eq:x-resonance}:
\be
\frac{\delta x_0} {\delta \lambda_{1n}} = \frac{d x_0}{d v} 
\frac{\delta v} {\delta \lambda_{1n}} =\left( \frac{v_0^{s-n}}{N_{s-n}}\right) P_{s-n}(v/v_0)\;. 
\ee
Therefore, the correlation is given as  
\be 
\langle O_{1,n} \rangle_D  
= 
\sqrt{ {\pi}{l}} \left(
 \frac{v_0^{s-n}}{N_{s-n}}\right) 
\int_{-v_0}^{-\infty} \dd v P_{s-n}(v/v_0)~  e^{  lv}
\sim \kappa^{s-n+1/2} K_{s-n+1/2} (l \kappa)\;,
\ee 
where the integration formula  is used: 
\be\label{Bessel-Legendre}
\int_1^{\infty} \dd x \; P_n(x) e^{-y x} = \sqrt{\frac{2}{\pi y}} ~K_{n+1/2}(y)\;. 
\ee 
The one-point correlations on the disk agree with  FZZ result~\cite{Fateev:2000ik}.

\subsection{M(3/5): new features}
\label{sec:4_M(3/5)}

The first non-trivial non-unitary example is M(3/5)  gravity,
which is based on $A_2$ Frobenius manifold. 
It is noted that  M(3/5) gravity has very different features  
from the unitary case.
For example, the variable $x$  vanishes at the lowest order 
of the perturbation. 
The integration path of $x$ cannot be chosen by putting $v_2=0$
since $x \propto v_2 $ vanishes identically in this case. 
As a result, the general framework breaks in M(3/5) gravity.  
The purpose of this section is to check carefully how one can cure 
the breakdown case by case. 
  
Douglas action of  M(3/5) is given with deformed parameters:
\begin{align}
S=   t_1^{(1,2)}  v_1 +  t_2^{(1,1)}  v_2-  H \;,~~~~
H=\theta_{2, 2} - t_{1,1}^{(13)} \theta_{1,1} 
- t_{1,2}^{(14)} \theta_{1,2}\;,\\
\theta_{2, 2}=- v_2 (v_1^3 - 3 v_2^2)/18  \;,~~~~~
\theta_{1,1} = v_1 v_2\;, ~~~~~
\theta_{1,2}= (-v_1^4 + 18 v_1 v_2^2)/36\;.
\end{align} 
The parameters have ${\mathbb Z}_2$ symmetry: 
$t_1^{(1,2)}=t_1^{(2,3)}$, 
$t_2^{(1,1)} = t^{(2,4)}$,
$t_{1,1}^{(13)} =t_{1,1}^{(22)}$,
$t_{1,2}^{(14)}=t_{1,2}^{(21)} $,
and g-dim is assigned as follows:
\be
[t_1^{(12) }]=7/6\;, ~~
[t_2^{(11)}]=1\;, ~~[t_{1,1}^{(13)}]=4/6\;, ~~[t_{1,2}^{(14)} ]=1/6\;.
\ee
The highest dimension is 7/6 and therefore $[x]= [t_1]> [\mu] $.
Instead, $[t_2^{(11)}]=1 =[\lambda_{11}] $
and $ \lambda_{11}=\mu$. 
According to g-dim analysis the following resonance transformations are allowed
\begin{align}
&t_1^{(12)} =\lambda_{12} + A^{(12)}_{(11),(14)} ~\mu   \lambda_{14}  
+ O(\lambda^4) \;,
&t_2^{(11)}= \lambda_{11}+ O(\lambda^3)\;, \\
&t_{1,1}^{(13)}  = \lambda_{13}  + O(\lambda^4)\;,
&t_{1,2}^{(14)} = \lambda_{14}\;.
\end{align}

The derivatives of the Douglas action have the form
\begin{align} 
\label{eq:35S1}
& \frac{\partial S}{\partial v_1}
 = t^{(12)}  -x; ~~~
x=- v_2  v_1^2 /6 
-t_{1,1}^{(13)}  v_2 
-t_{1,2}^{(14)}(-v_1^3/9+ v_2^2/2 )  \;, 
\\ 
\label{eq:35S2}
& \frac{\partial S}{\partial v_2} 
=t_2^{(11)} +  (v_1^3 - 9 v_2^2)/18  
 + t_{1,1}^{(13)} v_1 
+ t_{1,2}^{(14)}  v_1 v_2\;.
\end{align}  
On-shell solution is $v_2^*=0 $ in \eqref{eq:35S1}
and $v_1^{*3} = -18 \mu$ in \eqref{eq:35S2},
which looks  the same as  in the unitary series.
However, on-shell solution   $x^*$ vanishes
unlike the unitary case.  

Bulk generating function on  the sphere is of the form 
\eqref{eq:A2-sphere}.
With on-shell solution 
the same as  in the unitary series,
one may check that the bulk one-point correlators 
$ \langle O_{12} \rangle = \langle O_{14} \rangle =0$ 
vanish since $v_2^*=0$. 
On the other hand, the gravitational dimension $[O_{13}]=2$ and 
$ \langle O_{13} \rangle  =0 $ ( mod $\mu^2$)
as  in M(3/4) case.
Orthogonality of two-point correlations   
$ \langle O_{12}O_{13} \rangle =0=  \langle O_{13}O_{14} \rangle  $
is fulfilled  due to $v_2^*=0$. 
On the other hand, $ \langle O_{12}O_{14} \rangle $ has additional contribution due to the resonance terms
\be
   \langle O_{12}O_{14} \rangle 
      =  \int_0^{v_1^*} dv_1  ( - v_1^3 /9  + A_{(11),(14)} ^ {(12)}  \mu  ) \; .
\ee
The orthogonality condition fixes the resonance coefficient: 
\be 
 A_{(11),(14)} ^ {(12)} \mu  = -v_1^{*3}  /36 \;.
 \ee
For  two-point correlations this requirement 
leads to the same results as in the unitary case:  
the resonance transformations  are given in terms of Jacobi polynomials.  

As in the Lee-Yang series,
we use the generating function on the disk
\be
\label{eq:generating-function-34}
{\cal W}(l) = \int_{t_1^{(12)}} dx_0 
\int_{i \RR}  dy    e^{  l (y^3 + t v_1 + v_2)} \;,
\ee
with $x_0 = - v_2  v_1^2 /6 $. 

One-point correlator $\langle O_{1,2} \rangle_D$ is easily computed and 
has the form similar to \eqref{eq:34D-11}
\be 
\langle O_{12} \rangle_D   
  =  -  \int_{i \RR } dy  e^{  l (y^3 +v_1^* y)}
  = \frac{2 v_{1}^{*1/2}}{3} K_{1/3}(2l (v_{1}^*/3)^{3/2})\;,
 \ee
where $v_2^*=0$ is used.  This demonstrates that 
 $\langle O_{1,2} \rangle_D  \propto \kappa^{1/3} K_{1/3}(l \kappa)$.
 
Other correlators require more detailed information 
about $x_0$ and perturbation effects due to $\lambda_{1k}$'s.  
$
\langle O_{11} \rangle_D  $ is obtained by the perturbation of $\lambda_{11}$: 
\be
\langle O_{11} \rangle_D  
 =  \int_0^{\infty} dx_0 ~ \int_{i \RR } dy  
  ~  l
  \left ( y \frac{\delta v_1}{\delta \lambda_{11}} +  \frac{\delta v_2}{\delta \lambda_{11}}\right)_0
  \left. e^{ l (y^3 +v_1 y +v_2)} \right|_{0^*}  .
\ee
The variation of $v_1$ and $v_2$ due to $\lambda_{11}$ perturbation 
can be found from the string equation \eqref{eq:35S2}:  
$ {\delta  v_1}/{\delta \lambda_{1,1}} = -(6 /v_1^2) $ and 
$ {\delta  v_2}/{\delta \lambda_{1,1}} = 0 $.
Therefore, we have 
\be
\langle O_{11} \rangle_D  
 =-   l \int_0^{\infty} dx_0 ~ (6 /v_1^2)  \int_{i \RR } dy  
  ~ y    \left. e^{ l (y^3 +v_1 y +v_2)} \right|_{0^*}. 
\ee
Next step is to find the path of $x_0$-integration. 
Suppose we try to keep $v_2=0$  as in the unitary case. 
Then, one has $dx_0 = - dv_1 ( v_1 v_2/3)$ 
and the integration has null value. 
This suggests that we cannot follow the general framework used in the unitary case. 

Instead, we may prescribe the integration path alternatively:
keep the contour  path $v_1 = v^*_1$ and $v_2 \in [0, \infty]$.
In this case,  using $d x = -d v_2(v_1^2/6) $  we have  
\be
\label{eq:O11-v2}
\langle O_{11} \rangle_D  
 =  \int^{-\infty}_0  dv_2   \left( -  \frac{v_1^{*2}}6 \right)
 \int_{R } dy 
~ l y \left( -\frac{6}{ v_1^{*2}}   \right)
 e^{ il (y^3 +v_1 ^*y +v_2)}\;.
\ee 
The result is proportional to 
$ \kappa^{2/3} K_{2/3}( l \kappa ) $ as expected. 
Noting that the integrand is not closed,
the contour integration depends on the path deformation.
One needs proper integration path 
to achieve the right expectation value.
Therefore, the choice of the integration contour looks 
an ad hoc prescription to get an expected answer.

It is not clear at this moment whether there exists 
any canonical choice of the integration path. 
We present further examples illustrating this problem.   
The similar phenomenon occurs when we consider $ \langle O_{13} \rangle_D $.
Varying the string equation we get
\be
\frac{\delta v_1}{\delta \lambda_{13}} = -6 /v_1\;, \;
\frac{ \delta v_2}{\delta \lambda_{13}} = -  6 v_2 /v_1^2\;.
\ee 
Taking the  same integration contour as in \eqref{eq:O11-v2} we get
\begin{align}
 \langle O_{13} \rangle_D  
  = &\int^{-\infty}_0  dv_2   \left( -  \frac{v_1^{*2}}6 \right)
  \int_{i \RR } dy ~  l   ~
  \left( 
 \left (-   \frac{6 }{v_1^{*2} } v_2\right) + 
  y \left( -\frac{6}{ v_1^*}  \right)
  \right)
 e^{ l (y^3 +v_1 ^*y +v_2)} 
 \nn\\
  =&
    \int_{i \RR } dy 
  \left(   \frac 1l  -  y v_1^*  \right)   
 e^{ l (y^3 +v_1 ^*y +v_2)} 
 \propto \kappa^{4/3} K_{4/3} (l \kappa)\;,
 \label{eq:35O13}
 \end{align}
 which leads to the desired answer. 
 
One point function  $\langle O_{14} \rangle_D $ 
represents more complicated case. 
Suppose we keep $v_2 = 0$. 
Then, the string equation shows that 
\be
\frac{\delta v_1}{\delta \lambda_{14}} =0\;, \;~~
\frac{ \delta v_2}{\delta \lambda_{14}} = \frac    {2 v_1^2}   {3}\;. 
\ee 
Then the one-point function vanishes since the measure is proportional to $v_2$. 

One possible way out is to interpret the generating function 
in a different way.
Instead of using the variable $x_0$ in the generating function \eqref{eq:generating-function-34},
we may use  variable $x$ in  defining the generating function  
\be
\label{eq:generating-function-34}
{\cal W}(l) = \int_{t_1^{(12)}} dx  
\int_{i \RR}  dy    e^{  l (y^3 + t v_1 + v_2)} 
\ee
and  allow the variation  of the measure $dx$ while assuming $Q$ remains fixed so that 
\be
\langle O_{14} \rangle_D  
= 
 \int_{v_1^*} ^\infty dv_1
\left(   \left(  -\frac{v_1} 3\right) 
\left(  \frac{\delta v_2}{\delta \lambda_{14} }\right)  +    \frac{v_1^2}3  
  \right) 
   \int_{ i \RR }  dy 
    e^{ l (y^3 +v_1  y  )} 
\nn
 + \frac{v_1^{*3}}{36}~\langle O_{1,2} \rangle_D\;,
\ee 
which would lead to 
\begin{align}
\langle O_{14} \rangle_D  
= &
 \int_{v_1^*} ^\infty dv_1 \int_{ i R }
 dy 
\left(   \left(  -\frac{v_1} 3\right) 
\left(  \frac{2 v_1 }3 \right)  + 
   \frac{v_1^2}3  
  \right) 
    e^{ l (y^3 +v_1  y  )} 
\nn
 + \frac{v_1^{*3}}{36}~\langle O_{1,2} \rangle_D
\nn\\
\propto  &
\kappa^{7/3} ~ K_{7/3} ( l  \kappa)\;,
\end{align}
in agreement with the FZZ result.

\section{Concluding remarks}
\label{sec:concl}

We use the matrix formalism to compute bulk correlations of minimal gravity 
on the sphere and on the disk
and compare the result with the results of the Liouville gravity approach. 
We use the sphere correlation generating function   proposed in \cite{Belavin:2013nba}
and describe the effect of the resonance transformations for computing bulk correlators in the presence of boundary.
We clarify some subtleties in the construction of  the generating function 
if one generalize the matrix framework to $M(q/p)$ gravity, 
based on $A_{q-1}$ Frobenius manifold with $q>2$. 
We demonstrate that the 
matrix approach and the Liouville field approach 
agree with each other for the one and two correlations on the sphere and
for the one-point correlation on the disk.  

Even though the framework of the matrix approach 
looks similar for the unitary and non-unitary series, 
the details of the calculation show some subtleties. 
In  the unitary series $M(q/q+1)$, continuous variable $x$ of the 
matrix model approach has the same gravitational scaling dimension 
as the cosmological constant $\mu$.
For the non-unitary series, the gravitational dimension of $x$ is 
greater than that of cosmological constant. 
On the sphere, however, this subtlety is hidden 
because  $x$ does not come explicitly in the expression for the generating function. 
On the other hand, the computation of the bulk correlation on the disk
requires knowledge of the specific dependence of $x$  on the flat coordinates according to
the Douglas string equation. 
We provide some details of unitary series in section~\ref{sec:3}
and of non-unitary series in section~\ref{sec:4}
and find the consistency of the formalism 
once the subtleties are well taken care of. 

It is noted that on-shell condition defined in sec.~\ref{sec:2_scaling}, which is equivalent to
a specific solution to the Douglas equation,
is enough to find the correlation on the disk for unitary series,
whose calculation is simplified 
in the flat coordinates of the Frobenius manifold. For the non-unitary Lee-Yang series $M(2/2s+1)$, the Frobenius manifold is one dimensional $A_1$ and hence is trivial. In this case, one can simply use the off-shell condition for the flat coordinate. 
However, for higher $q$ non-unitary models, when the Frobenius manifold becomes multi-dimensional, the usage of the matrix model results requires more profound analysis of the off-shell condition. The considered examples show that 
there is no canonical way of evaluating the one-point correlation on the disk 
for the non-unitary models without using some additional assumptions. 
Even though there are certain conjectures about the choice of the  contours and the off-shell application of the string equation 
leading to the desired FZZ result, 
the analytic structure of 
the generating function on the disk  is not fully understood   
and needs further investigation. 

Among other possible further developments of the present analysis we mention the following natural questions.
The case of non-trivial boundary conditions as well as of the other types of correlations involving boundary insertions remains to be studied. 
In particular, the resonance transformations  in this case 
can be affected by the presence of boundary fields and further analysis of the Frobenius manifold structure  hidden beyond the 
matrix model formulation is required. In this context the analysis of the connection between the
intersection theory on the moduli space of Riemann surfaces with boundary and the open KdV equations might be useful (see, e.g. \cite{Pandharipande:2014qya,Buryak:2014apa,Buryak:2014dta,Bertola:2014yka}). Finally, we note that the new methods of the dual approach to the minimal (open and closed) strings based
on the connection with the theory of Frobenius manifolds suggests that maybe some of these results can be 
applied for analytic computations of string amplitudes in more realistic string models. 
In this respect the consideration in \cite{Balthazar:2017mxh}, where $c = 1$ string theory has been considered from the worldsheet perspective and compared   with the matrix models, may be relevant.

\vspace{7mm}
\noindent \textbf{Acknowledgements.} The work of K.A. and V.B. was supported by the Foundation
for the advancement of theoretical physics ``BASIS''. The work of C.R. was partially supported by National Research Foundation of Korea grant NRF-2017R1A2A2A05001164.

\appendix

\section{Bessel functions and Chebyshev polynomials} \label{sec:app}

Below we list some useful formulae for Bessel
K-functions and Chebyshev polynomials

\bal
& K_{\nu}(z) = \int_0^{\infty} \dd s \: \ch(\nu s) \: e^{-z \ch(s)}\;, \\
& K_{\nu}(z) = \frac{1}{\co(\nu\pi/2)}\int_0^{\infty} \dd s \: \ch(\nu s) \: \co (z \sh(s))\;, \\
& K_{\nu}(z) = \frac{1}{\si(\nu\pi/2)}\int_0^{\infty} \dd s \: \sh(\nu s) \: \si (z \sh(s))\;, \\
& T_n(i \sh(s)) = (-1)^{(n-1)/2} i \: \sh(n s)\;, \; n = 2k-1\; , \\
& T_n(i \sh(s)) = (-1)^{n/2} \ch(n s)\;, \; n = 2k \; .
\eal
To perform the $x$ integration we use:
\bal
&\int_{\mu}^{\infty} \dd x \, x^{\nu/2} K_{\nu}(x^{1/2}) = 2\mu^{(\nu+1)/2} K_{\nu+1}(\mu^{1/2})\;, \\ 
&K_{\nu}(x) = K_{-\nu}(x)\;, \\
&K_{\nu}(x) = \frac{x}{2\nu}\left(K_{\nu+1}(x) - K_{\nu-1}(x) \right)\;.
\eal
In particular, for $M(3,p)$ models we have
\be
\int_{\RR} \dd y \; e^{il(y^3 + v_1 y)} = \frac{2 v_1^{1/2}}{3} K_{1/3}(2l (v_1/3)^{3/2})\;.
\ee


 \bibliographystyle{JHEP}
 \bibliography{mlg1}

\end{document}